\documentclass[twocolumn,nofootinbib,floats,amsmath,amssymb]{revtex4}
\usepackage{graphicx}
\usepackage{amsmath}
\usepackage{float}
\usepackage{amssymb}
\usepackage[all]{xy}
\usepackage{amsfonts}
\usepackage{dcolumn}
\usepackage{bm}
\usepackage{xcolor}

\begin{document}

\title{Challenging matter creation models in the phantom divide}
 
\author{V\'ictor H. C\'ardenas$^1$}
\email{victor.cardenas@uv.cl}

\author{Miguel Cruz$^2$}
\email{miguelcruz02@uv.mx}

\author{Samuel Lepe$^3$}
\email{samuel.lepe@pucv.cl}

\author{Shin'ichi Nojiri$^{4,5}$}
\email{nojiri@gravity.phys.nagoya-u.ac.jp}

\author{Sergei D. Odintsov$^{6,7}$}
\email{odintsov@ieec.uab.es, odintsov@ice.csic.es}

\affiliation{$^1$Instituto de F\'{\i}sica y Astronom\'ia, Universidad de Valpara\'iso, Gran Breta\~na 1111, Valpara\'iso, Chile\\
$^2$Facultad de F\'{\i}sica, Universidad Veracruzana 91000, Xalapa, Veracruz, M\'exico\\
$^3$Instituto de F\'{\i}sica, Facultad de Ciencias, Pontificia Universidad Cat\'olica de Valpara\'\i so, Avenida Brasil 2950, Valpara\'iso, Chile\\
$^4$Department of Physics, Nagoya University, Nagoya 464-8602, Japan\\ 
$^5$Kobayashi-Maskawa Institute for the Origin of Particles and the Universe, Nagoya University, Nagoya 464-8602, Japan\\
$^{6}$Institut de Ci\`encies de l'Espai, ICE/CSIC-IEEC, Campus UAB, Carrer de Can Magrans s/n, 08193 Bellaterra (Barcelona), Spain\\
$^{7}$Instituci\'o Catalana de Recerca i Estudis Avan\c{c}ats (ICREA), Barcelona, Spain}

\date{\today}

\begin{abstract}
We perform a study both statistical and theoretical for cosmological models of matter creation and their ability to describe effective phantom models of dark energy. Such models are beyond the $\Lambda$CDM model since the resulting cosmic expansion is not adiabatic. In fact, we show that this approach exhibits transient phantom/quintessence scenarios at present time and tends to the standard cosmological model at some stage of the cosmic evolution. We discuss some generalities of the  thermodynamics properties for this type of cosmological model; we emphasize on the behavior of the temperature associated to dark matter, which keeps positive along cosmic evolution together with the entropy. The enrichment of this type of model by means of the incorporation of cosmological constant and dissipative effects in the fluid description to explore their cosmological consequences in the expansion of the universe is considered. Finally, a generalization for the matter production rate as an inhomogeneous expression of the Hubble parameter and its derivatives is discussed; as in all the cases examined, such election leads to an effective phantom/quintessence behavior. 
\end{abstract}


\maketitle

\section{Introduction} 

Since it was discovered that supernovae were dimmer than expected, we have not been able to establish the reason for this behavior rather than assume the simplest: the existence of a global cosmological constant. Although the discovery of this behavior occurred more than 20 years ago, and during this time the large scale observations have given us an accuracy never reached before, we have not been capable to elucidate the cause of this behavior. Assuming that observations are correct, at theoretical level we only have three options: either to assume the existence of a new component of the universe, one whose pressure is negative, or also to assume that we must modify the theory of gravitation, or the third option, assume that our universe at the scales of measurements does not satisfy the Copernican principle.

Within the first category, and assuming that the extra component -- usually called dark energy (DE) -- satisfies an equation of state $p = \omega_{\mathrm{de}} \rho$, with $\omega_{\mathrm{de}}$ constant, the observations are used to constrain the best fit value of $\omega$ among other cosmological parameters. However, many recent observations \cite{obs, plancknew} indicate that $\omega_{\mathrm{de}} < -1$. If we interpret $\omega_{\mathrm{de}}$ as the EoS parameter of a single component, several physical complications appear. First, the condition $\omega_{\mathrm{de}} < -1$ can not be achieved by Einstein gravity itself; secondly, if we assume the existence of a fundamental scalar field (sometimes called phantom matter) that satisfies the aforementioned condition for the EoS parameter, we must deal with a non canonical Lagrangian, i.e., the kinetic term carries a negative sign. This implies that the dominant energy condition (DEC) is not satisfied, in consequence, propagation of energy outside the light cone and vacuum instabilities are expected to appear in this kind of model \cite{dec1, dec2}. In Ref.~\cite{odintsovcft} can be found that it is possible to preserve the energy conditions for an accelerating universe with phantom and ordinary matter, but the model requires quantum effects contribution in the phantom sector. On the other hand, if the phantom scalar field is coupled to a perfect fluid, the r.h.s. of Einstein equations can be written as the sum, $T_{\mu \nu} := T^{\mathrm{fluid}}_{\mu \nu} + T^{\mathrm{ph}}_{\mu \nu}$. By means of the Bianchi identity $T_{\mu \nu}$ is conserved but each term of the sum not; a simple interpretation of this is that matter is being created by the phantom field in order to maintain a constant matter density. However, another consequence from the conservation condition, $\nabla^{\mu}T_{\mu \nu}$, is that the rate of enthalpy production acts as a source for the phantom field, this is contradictory since in the standard scheme the entropy is a constant times the enthalpy density and such entropy is constant \cite{gibbons}.\\ 

As can be seen, although the phantom field approach may be in accordance with the observations, it has several problems at fundamental level, such as those that we have already mentioned above and some other characteristics not well seen as a future singularity as final fate for the universe \cite{prl}; a critical problem in this context is the definition of physical quantities in the neighborhood of such singularity. However, it can be found that again the quantum gravity effects can help to keep some of these quantities well defined near the singularity, see for example the Refs.~\cite{cft1, cft2}.\\         

One way to overcome some of the aforementioned problems of the phantom field, is to imagine that maybe $\omega_{\mathrm{de}}<-1$ is an {\it effective} result. In Ref.~\cite{Nunes:2015rea} the authors suggested a model where such result is possible if we consider the simultaneous action coming from $\Lambda$ (with $\omega_{\Lambda}=-1$) plus the negative contribution from matter creation. We will focus on this approach but it is worthy to mention that is not the only possibility to avoid the phantom difficulties, in Ref.~\cite{Nojiri:2005sr} was found that a phantom scenario was possible to obtain by generalizing the EoS of the cosmological fluid to an inhomogeneous expression of the Hubble parameter with no need of introducing negative energy considerations. As we will discuss later, an adequate description of the universe at late times can be obtained in the approach of matter creation models if one considers an inhomogeneous generalization for the matter production rate expression. Models of matter creation can be found in the literature and have been shown that could play a relevant role in the early universe \cite{lima} or on the consistency of the thermodynamics description of some generalizations of Einstein gravity \cite{navarrete}.\\

In this work we study some models of creation of particles and their possibilities of realizing an effective phantom component alone, with and without adding a cosmological constant. We also study some thermodynamic properties of these systems, we focus on the temperature of the created matter, this is definite positive and its behavior seems to be in agreement with some recent results, we also consider the introduction of other effects in the cosmological fluid to explore their cosmological consequences. The work is organized as follows. In Section \ref{sec:nunes} we briefly discuss the ideas given by the authors of Ref.~\cite{Nunes:2015rea}. In Section \ref{sec:close} we explore two possibilities for the matter creation rate and we show that in each case the phantom (quintessence) scenario is possible to achieve at present time but this is only transient, the models tend to evolve to a de Sitter expansion. In Section \ref{sec:temp} we consider several possibilities for the matter creation rate in order to discern if we could have different thermodynamics scenarios. In Section \ref{sec:observations} we perform the statistical analysis of the model discussed in Section \ref{sec:close}. In section \ref{sec:inhomo} we consider a generalization of the matter production rate as a function of the Hubble parameter and also of its derivatives. Finally, in Section \ref{sec:final} we give the final comments of our work.

\section{Matter creation plus lambda as phantom}
\label{sec:nunes}

Following the line of reasoning of Ref.~\cite{Nunes:2015rea}, if matter creation exists, i.e., gravitational particle production, then for a FLRW spacetime $\Gamma \neq 0$, yielding
\begin{equation}\label{conseqs}
\dot{n} + 3H n = n \Gamma, \hspace{1cm} \dot{\rho} + 3H (\rho + P) = 0,
\end{equation}
where $\Gamma >0$, $\Gamma < 0$ acts like a source or sink of particles, respectively; $n$ is the particle number density and $P=p+\Pi$. Here $\Pi$ accounts for the pressure from matter creation (sometimes written as $p_c$). From the Gibbs equation
\begin{equation}\label{gibbseq}
T dS = d \left(\frac{\rho}{n}\right) + p d \left(\frac{1}{n}\right),
\end{equation}
we can write
\begin{equation}
    nT\dot{S}=-3 H \Pi - (\rho + p)\Gamma,
\end{equation}
If we assume $\dot{S}=0$, i.e., the case of adiabatic particle creation we have
\begin{equation}
    \Pi = - \frac{\rho+p}{3H} \Gamma,
    \label{eq:pi}
\end{equation}
then as the authors said, the effective EoS parameter (assuming $p=0$) is the sum of the EoS of vacuum plus the contribution due to dark matter (DM) creation
\begin{equation}
    \omega_{\mathrm{eff}}=\frac{p_{\Lambda}}{\rho_{\Lambda}}+\frac{p_{c}}{\rho_{\mathrm{dm}}} = -1-\frac{\Gamma}{3H},
\end{equation}
showing that is possible to obtain a $\omega_{\mathrm{eff}}<-1$. Let us revise the arguments carefully. The cosmological model of Ref.~\cite{Nunes:2015rea} consists in DM plus a cosmological constant, i.e.,
\begin{equation}\label{sys0}
     \dot{\rho}_{\mathrm{dm}} + 3H (\rho_{\mathrm{dm}} + \Pi) = 0,\hspace{1cm} \dot{\rho_{\Lambda}}=0,
\end{equation}
where here the DM already incorporates the gravitational matter production pressure $\Pi$, and we have assumed already that $p_{\mathrm{dm}}=0$ and $p_{\Lambda}=-\rho_{\Lambda}$. We would like to stress here that the observational result $\omega_{\mathrm{de}}<-1$ for DE is obtained {\it together} with the assumption of a non-relativistic DM component contribution evolving as $\tilde{\rho}_{\mathrm{dm}} \propto a^{-3}$, i.e., with $\tilde{\omega}_{\mathrm{dm}}=0$. In fact, we are ``measuring'' a model described by
\begin{equation}\label{sys1}
     \dot{\tilde{\rho}}_{\mathrm{dm}} + 3H \tilde{\rho}_{\mathrm{dm}} = 0,\hspace{0.5cm} \dot{\rho_x}+3H(1+\omega)\rho_x=0,
\end{equation}
where observationally $\omega < -1$ (where it is clear that $p_x = \omega \rho_x$). The question now is: Is it possible to confuse (\ref{sys0}) with a phantom cosmology described by (\ref{sys1})? Because the observables depends directly on the Hubble function $H(a)$, in both cases this expression {\it must be the same}, so because $H^2 \propto \rho_{\mathrm{tot}}$ the invariant quantity is
\begin{equation}\label{rhotot}
    \rho_{\mathrm{tot}} = \rho_{\mathrm{dm}} + \rho_{\Lambda} = \tilde{\rho}_{\mathrm{dm}} + \rho_x.
\end{equation}
As we will see below, our results differ from those discussed in \cite{Nunes:2015rea}.

\section{Models of effective phantom using only matter creation: exploring solutions close to $\Lambda$CDM}
\label{sec:close}
Let us consider a couple of specific solutions of gravitational DM production model. In this case we refer as DM as the non-relativistic fluid characterized by $\rho_{\mathrm{dm}} = m n$ where $m$ is the mass of the particle and $n$ is the number density satisfying (\ref{conseqs}). By using the model $n\Gamma=3H\alpha$ \cite{victor, victor1} we get that
\begin{equation}
n(a)= \frac{n_0 - \alpha}{a^{3}} + \alpha,
\end{equation}
that implies a solution $\rho_{\mathrm{dm}}$ that resembles the combined contribution of a dust component ($\propto a^{-3}$) plus a constant energy density, mimicking in this way the $\Lambda$CDM model. For this particular interaction function, the model does not permit to cross the phantom line. The important thing obtained here is that it is possible to get an universe which is similar to the $\Lambda$CDM model without adding an {\it ad hoc} negative pressure contribution (or cosmological constant), we just need to allow gravitational production of DM. It is also naturally resolved the coincidence problem, basically because both contributions (dust plus cosmological constant) are produced by the same source. 

\subsection{The $\Gamma$ constant case}

\begin{figure}[htbp!]
\centering
\includegraphics[width=7.8cm,height=6cm]{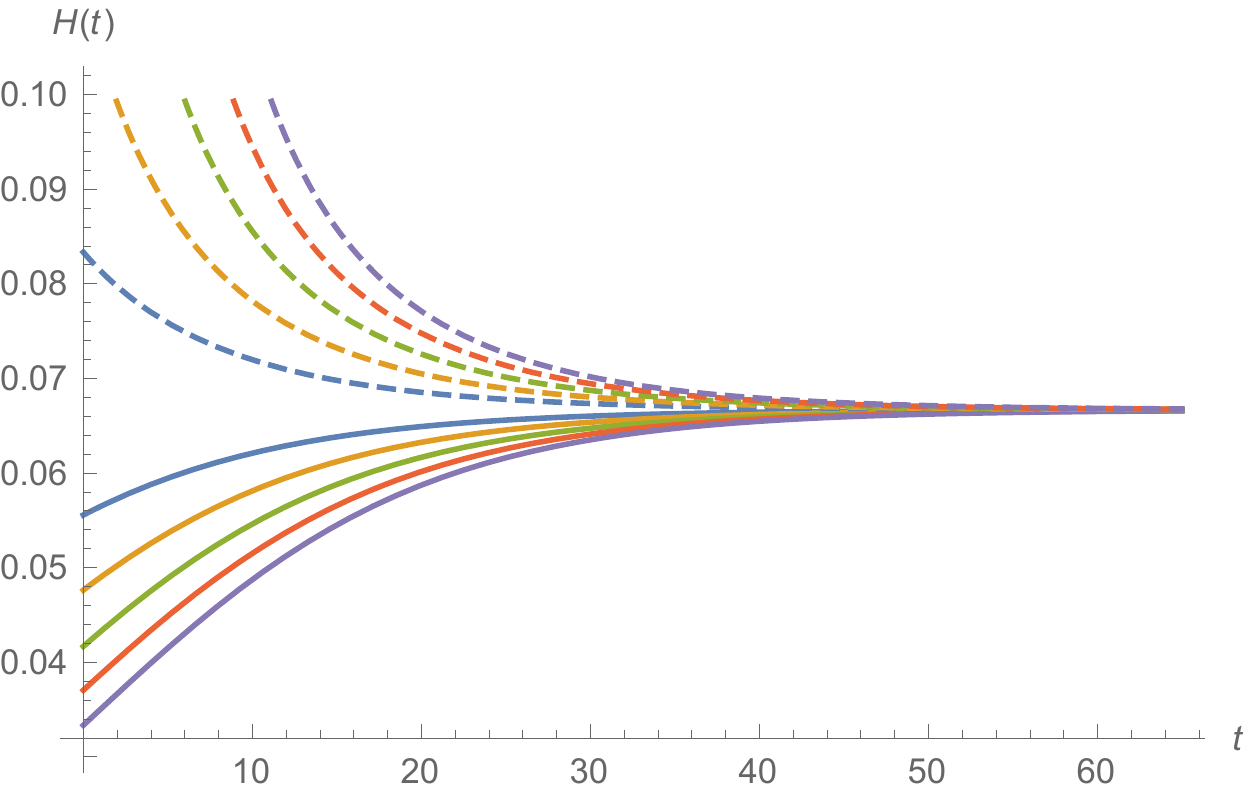}   
\caption{Hubble parameter. The solid lines represent the case $A > 0$ and the dashed lines correspond to $A < 0$. For all plots we have considered $\Gamma = 0.2$ and besides $A = 0.2,  0.4, 0.6, 0.8, 1$ (from upper to lower solid lines) and $A = -0.2,  -0.4, -0.6, -0.8, -1$ (from lower to upper dashed lines), as observed, the initial value for the Hubble parameter increases as $A$ approaches to $-1$.}
\label{fig:hubble}
\end{figure}

\begin{figure}[htbp!]
\centering
\includegraphics[width=7.8cm,height=6cm]{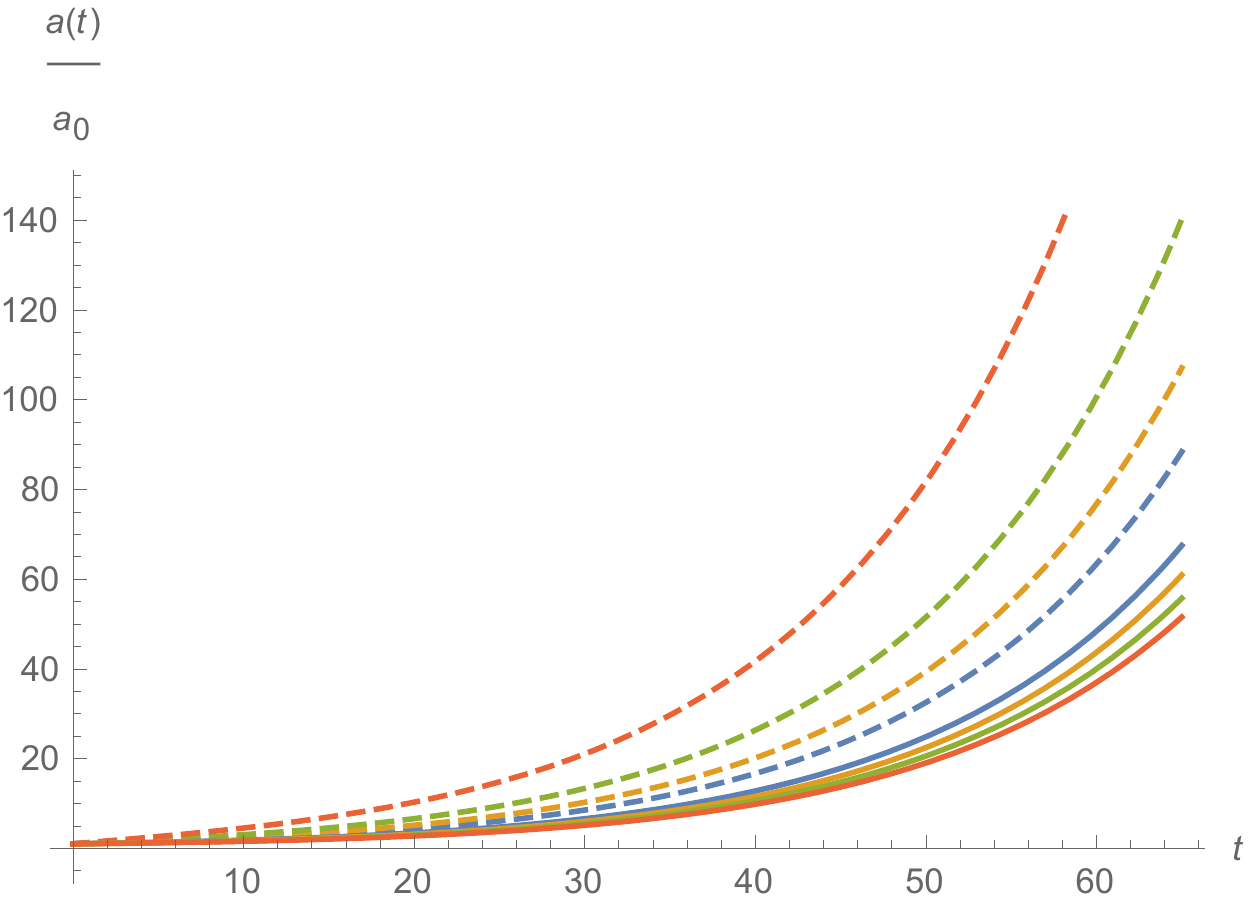}   
\caption{Scale factor evolution. For all plots we have considered $\Gamma = 0.2$ and $A = 0.2, 0.4, 0.6, 0.8$ (from upper to lower solid lines) and $A = -0.2, -0.4, -0.6, -0.8$ (from lower to upper dashed lines).}
\label{fig:scale}
\end{figure}

Now, let us assume an interaction model with $\Gamma=\text{constant}$. In this case from (\ref{conseqs}) we get 
\begin{equation}\label{romsol}
    \rho_{\mathrm{dm}} =  \frac{\rho_{\mathrm{dm,0}}}{a^3}e^{\Gamma \Delta t},
\end{equation}
recall that $\rho_{\mathrm{dm}} = m n$. From the Friedman equation we know that, $3H^2 = \rho$, if we also consider the continuity equation (\ref{conseqs}) for the energy density, we can write the following differential equation for the Hubble parameter
\begin{equation}\label{hdteq}
    \frac{\dot{H}}{H^{2}} = -\frac{3}{2}\left(1-\frac{\Gamma}{3H}\right).
\end{equation}
Because $\Gamma$ is constant we find that
\begin{equation}\label{hdasol}
    H(a) = \frac{\Gamma}{3} + \frac{H_0}{a^{3/2}} \left(1 - \frac{\Gamma}{3H_0} \right)
\end{equation}
We can also solve the equation (\ref{hdteq}) where time is explicit in the solution. Using that $H=\dot{a}/a$ we get
\begin{equation}\label{hdtsol}
    H(t)=H_0 a^{-3/2} \exp(\Gamma (t-t_0)/2).
\end{equation}
Combining (\ref{hdasol}) and (\ref{hdtsol}) we get
\begin{equation}
    \exp{(\Gamma \Delta t/2)} = (A+1)a^{3/2} - A,
\end{equation}
where we have defined $A = \Gamma/3H_0 - 1$. Then replacing this in (\ref{romsol}) we get
\begin{equation}
    \rho_{\mathrm{dm}} (a) = \frac{\rho_{\mathrm{dm,0}}}{a^3} \left((A+1)a^{3/2} - A \right)^2.
    \label{eq:density}
\end{equation}
As is evident, in this last expression we can recognize three contributions: a dust like evolving as $a^{-3}$, a component that evolves as a fluid with EoS parameter $\omega = -1/2$, and a component that evolves as a cosmological constant. From (\ref{hdteq}) we can write an effective expression for the pressure, that results in
\begin{equation}
    p_{\mathrm{eff}}=-\Gamma H = \frac{\Gamma^2}{3} - A \frac{H_0 \Gamma}{a^{3/2}}.
\end{equation}
If $A>0$, then we have the possibility of getting a phantom at effective level. According to Eq. (\ref{hdteq}) we can write the Hubble parameter as an explicit function of time as follows
\begin{equation}
H(t) = \frac{\Gamma}{3}\left[1+A \exp \left(-\frac{1}{2}\Gamma(t-t_{0}) \right) \right]^{-1},
\label{eq:hubb}
\end{equation}
and from the previous expression we get for the scale factor
\begin{equation}
a(t) = a_{0}\left[\frac{1+(1/A)\exp\left(\frac{\Gamma}{2}(t-t_{0})\right)}{1+(1/A)} \right]^{2/3}.  
\label{eq:scale}
\end{equation}
For an expanding universe we must have $H>0$, therefore from the expression (\ref{eq:hubb}) we can see straightforwardly that the condition $A > 0$ must be satisfied, i.e., $\Gamma > 3H_{0}$ and besides $H(t=t_{0}) = \Gamma/3(1+A) > 0$, the initial value for the Hubble parameter depends only on the values of $\Gamma$ and $A$, notice that the positivity for this initial value can be guaranteed always that, $A > -1$, therefore the model admits a region in which we could have $A < 0$. In Fig. (\ref{fig:hubble}) we show the behavior of Eq. (\ref{eq:hubb}) by considering a fixed value for $\Gamma$ and varying the value of $A$. As observed, the Hubble parameter starts from an initial value and tends to a constant value as time increases, which is given by $\Gamma/3$ and according to the values used in the plots is around $0.066$. This represents a similarity with the $\Lambda$CDM model for the cosmic evolution, this can be seen from Eq. (\ref{eq:density}), as universe expands the leading term in the energy density is given by the constant term $\rho_{\mathrm{dm,0}}(1+A)^2$.
Using the same values for $\Gamma$ and $A$ as in the previous plot, in Fig. (\ref{fig:scale}) we show the behavior of the quotient, $a(t)/a_{0}$, given by Eq. (\ref{eq:scale}). As can be seen in all cases we have an initial value equal to $1$ for the quotient and grows as time increases. For $A < 0$ we can observe that such quotient grows faster.
From Eq. (\ref{eq:hubb}), after a straightforward calculation we can obtain the following expression for the deceleration parameter 
\begin{equation}
q(t) = -1-\frac{3}{2}A \exp \left[-\frac{\Gamma}{2}(t-t_{0}) \right],    
\end{equation}
where we have, $q(t \rightarrow \infty) \rightarrow -1$, which represents a cosmological constant evolution and $q(t=t_{0}) = -1-3A/2 < -1$ for $A >0$. On the other hand, for $-1 < A < 0$ we have, $-1 <  q(t=t_{0}) < 1/2$, i.e., at present time this universe could have a phantom-like evolution or a quintessence behavior, depending on the value of the constant $A$.\\ 

Alternatively, the continuity equation (\ref{conseqs}) for the energy density can be written as
\begin{equation}
\dot{\rho} + 3H(1+\omega_{\mathrm{eff}})\rho = 0,
\label{eq:omegeff}
\end{equation}
where we have defined the effective parameter state
\begin{equation}
\omega_{\mathrm{eff}} = -\frac{\Gamma}{3H},
\label{goodone}
\end{equation}
since we have considered the pressure from matter creation as given in Eq. (\ref{eq:pi}) with $p=0$ and a barotropic EoS between the energy density and the aforementioned pressure. Note that the expression for the effective parameter depends strongly on the particle production rate, $\Gamma$, once we define it, the differential equation (\ref{hdteq}) for the Hubble parameter becomes solvable. Therefore, we will consider as effective parameter state the expression given in (\ref{goodone}) for any matter creation model.\\

\begin{figure}[htbp!]
\centering
\includegraphics[width=7.8cm,height=6cm]{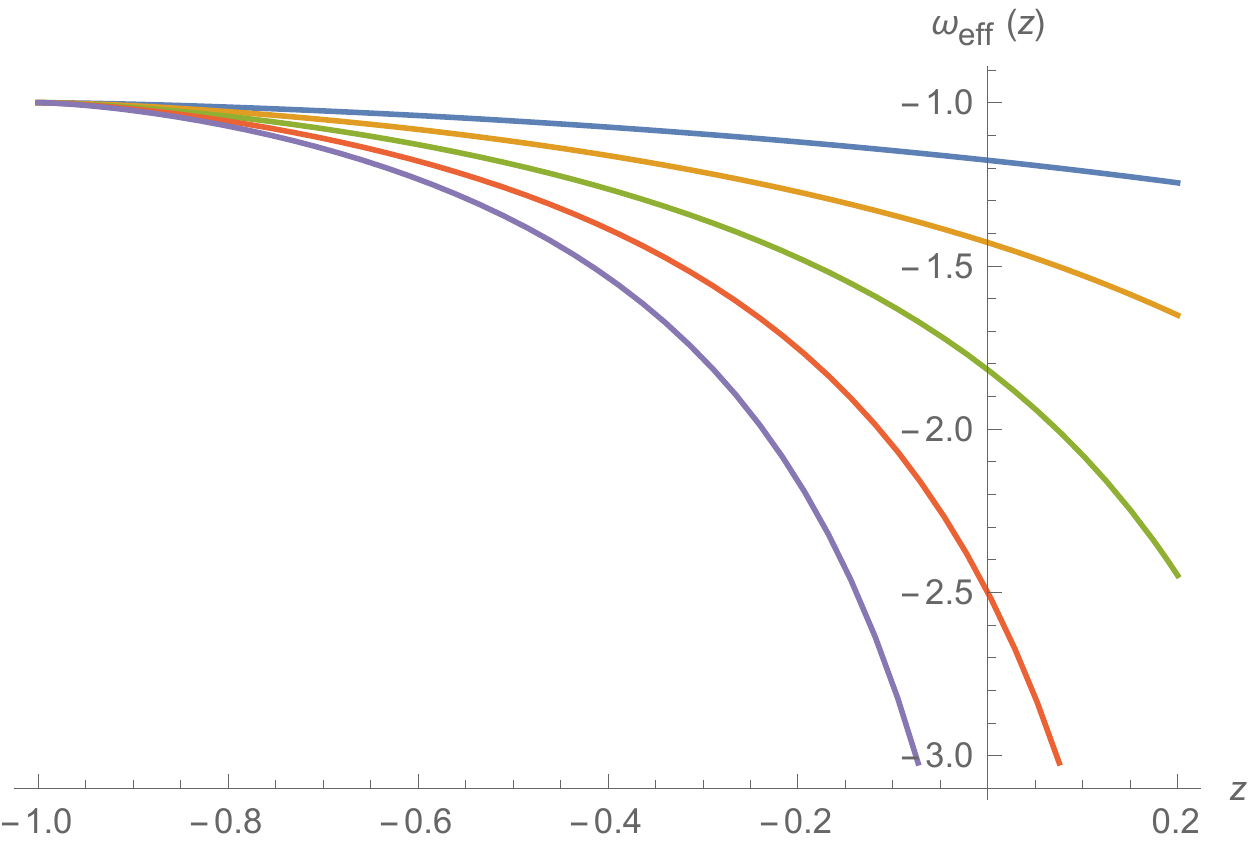}
\includegraphics[width=7.8cm,height=6cm]{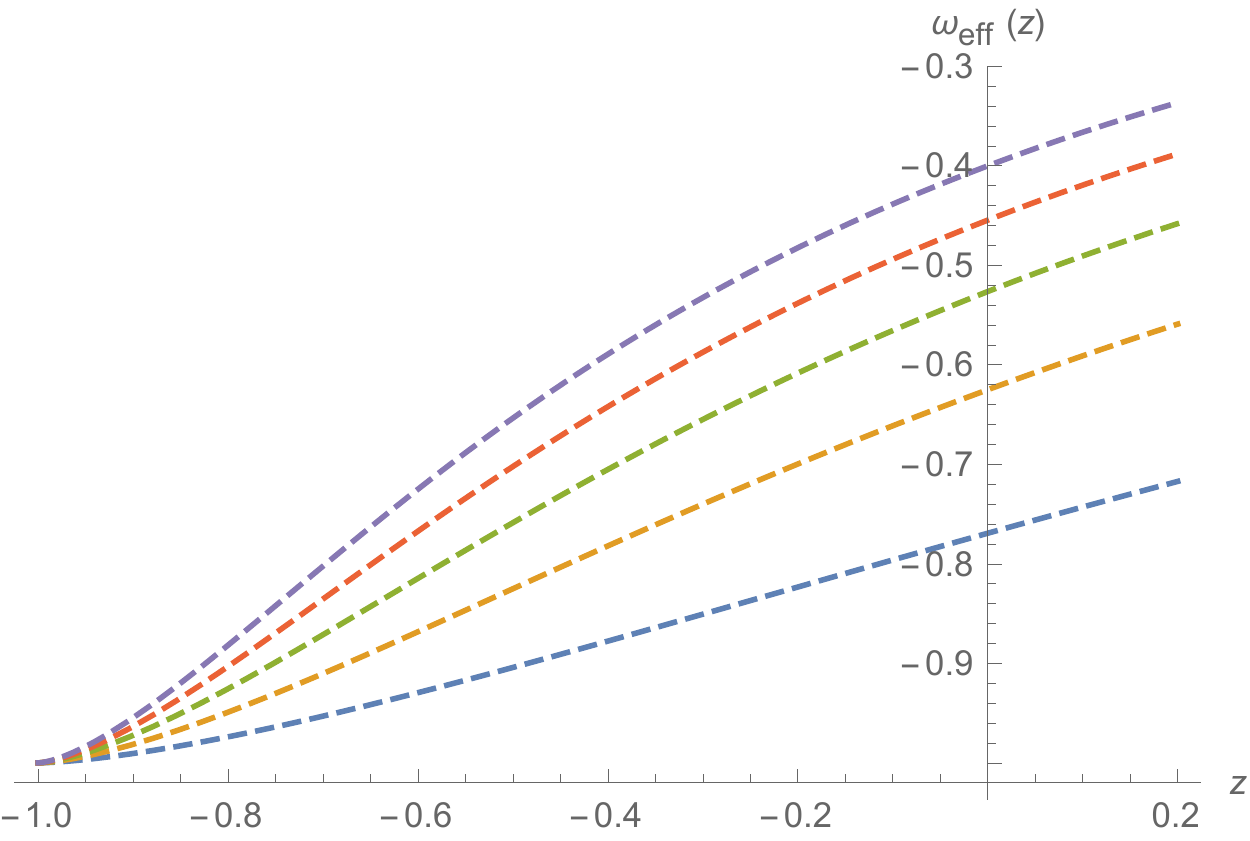}
\caption{Effective parameter state as a function of the redshift. In the upper panel we have considered $A = 0.2,  0.4, 0.6, 0.8, 1$ (from upper to lower solid line) and $A = -0.2,  -0.4, -0.6, -0.8, -1$ (from lower to upper dashed line).}
\label{fig:omega}
\end{figure}

Using the expression of the Hubble parameter given in Eq. (\ref{hdasol}) and the conventional relation between the redshift and the scale factor, $1+z = a^{-1}$, we have for the effective parameter state
\begin{equation}\label{weff}
\omega_{\mathrm{eff}}(z) = -\frac{\Gamma}{\Gamma-3H_{0}(1+z)^{3/2}A}.      
\end{equation}
In Fig. (\ref{fig:omega}) we illustrate the behavior of this effective parameter state taking into account both situations, the upper panel shows the case $A>0 \ (\Gamma > 3H_{0})$ and the case $A<0 \ (\Gamma < 3H_{0})$ it is shown in the lower panel, for simplicity we have considered the value of $H_{0}$ equal to one. As shown in the upper panel of the plot, the cosmological model evolves from an over accelerated stage ($\omega_{\mathrm{eff}} < -1$) to a cosmological constant evolution ($\omega_{\mathrm{eff}} = -1$) as we approach to the far future, therefore in order to obtain an accelerated expansion there is no need of introducing some extra components such as DE when DM production is considered; this result is in agreement with Eq. (\ref{eq:density}), where at some stage of cosmic evolution the leading terms are given by a dust like component and a fluid with parameter state $\omega = -1/2$, and as time evolves the leading term is simply given by a constant. On the other hand, the lower panel of the plot shows that the model evolves from quintessence to a cosmological constant expansion. Given that no other fluid is introduced in this picture, this kind of model is reasonable to describe only the late times evolution.

As can be seen, in both cases the phantom-like (quintessence) behavior is only a transient stage of the model, being the final fate for these universes a de Sitter expansion. This peculiar feature was also obtained in the context of the DGP braneworld model in Ref.~\cite{transient1} and also in Ref.~\cite{transient2} in the General Relativity framework, where was found that the backreaction due to the particle production is capable to stabilize an universe dominated by a phantom component.

\subsection{$\Gamma \propto H^{\alpha}$}

In this section we will consider the following general expression for the particle production rate
\begin{equation}
\Gamma = 3\beta H_{0}\left(\frac{H}{H_{0}} \right)^{\alpha},     
\end{equation}
where $\alpha$ and $\beta$ are dimensionless constants and $H_{0}$ represents the Hubble constant. If we insert the previous expression in Eq. (\ref{hdteq}) and consider the redshift definition given previously, the differential equation for the Hubble parameter becomes
\begin{equation}
    (1+z)\frac{dH}{dz} = \frac{3}{2}H_{0}\left\lbrace \frac{H}{H_{0}}-\beta \left(\frac{H}{H_{0}} \right)^{\alpha} \right\rbrace,
\end{equation}
which solution is given as follows
\begin{eqnarray}
E\left( z\right) =\left\{ 
\begin{array}{c}
\left[ \beta +\left( 1-\beta \right) \left( 1+z\right) ^{3\left( 1-\alpha
\right) /2}\right] ^{\frac{1}{1-\alpha }}, \ \alpha \neq 1,
\\ 
\left( 1+z\right) ^{3\left( 1-\beta \right) /2}, \ \ \alpha
= 1,
\label{besth}
\end{array}
\right. 
\end{eqnarray}
where, $E(z) := H(z)/H_{0}$, and it is usually termed as normalized Hubble parameter. This solution was studied in Refs.~\cite{generalrate, generalrate1} at background and perturbative levels. Therefore, the deceleration parameter, $1+q = - \dot{H}/H^{2} = (1+z)d \ln E(z)/dz$, takes the form
\begin{equation}
q\left( z\right) = -1+\frac{3}{2}\frac{(1-\beta)(1+z)^{3(1-\alpha)/2}}{\beta + (1-\beta)(1+z)^{3(1-\alpha)/2}}, \ \alpha \neq 1,
\end{equation}
for the case $\alpha = 1$, the deceleration parameter becomes a constant. From the previous expression we can observe that for $\alpha < 1$, $q(z\rightarrow -1) \rightarrow -1$, i.e., as we approach the far future we obtain a cosmological constant evolution; this result is independent of the value of the constant $\beta$. At present time, $q(z=0) =(1-3\beta)/2$, if $\beta > 1/3$ we have a $q(0) < 0$. On the other hand, the effective parameter state (\ref{goodone}) will be given by 
\begin{equation}
\omega_{\mathrm{eff}}(z) = -\beta (E(z))^{\alpha-1},
\label{bestomega}
\end{equation}
and at present time we obtain, $\omega_{\mathrm{eff,0}} := \omega_{\mathrm{eff}}(z=0) = -\beta$, which is independent of the constant $\alpha$, given that the condition $\beta > 1/3$ must be satisfied to have an accelerated stage at present time; therefore the model could have transient quintessence or phantom scenarios. Additionally, for $\alpha > 1$ we have, $q(z\rightarrow -1) > 0$, independently of the value of the constant $\beta$, therefore this case is not considered for analysis.

\section{Dark matter temperature}
\label{sec:temp}

Since we are considering the particle number density, $n$, we must modify the first law as follows \cite{victor}
\begin{equation}
    d(pV)+pdV-\frac{\rho+p}{n}d(nV)=0,
\end{equation}
therefore by taking its time derivative we can write for the energy density
\begin{equation}
    \dot{\rho}=\frac{\dot{n}}{n}(\rho+p),
    \label{eq:contmod}
\end{equation}
note that previous equation takes its standard form when $\Gamma = 0$, from Eq. (\ref{conseqs}) for the density number we can see that, $\dot{n}/n = -3H$, for $\Gamma = 0$. On the other hand, since the temperature is defined by the Gibbs equation given in (\ref{gibbseq}) we have, $T = T(n, \rho)$ \cite{maartens}, then
\begin{equation}
    \dot{T} = \frac{\partial T}{\partial n}\dot{n} + \frac{\partial T}{\partial \rho}\dot{\rho} = \frac{\dot{n}}{n}\left[\frac{\partial T}{\partial n}n + \frac{\partial T}{\partial \rho}(\rho + p)\right],
    \label{eq:dott}
\end{equation}
where we have considered the Eq. (\ref{eq:contmod}). Using the integrability condition, $\partial^{2}S/\partial T \partial n = \partial^{2}S/\partial n \partial T$, one gets 
\begin{equation}
    \frac{\partial T}{\partial n}n + \frac{\partial T}{\partial \rho}(\rho + p) = T\frac{\partial p}{\partial \rho},
\end{equation}
and using the previous result in Eq. (\ref{eq:dott}) we obtain the evolution equation for the temperature given as follows
\begin{equation}
    \frac{\dot{T}}{T} = \frac{\dot{n}}{n}\frac{\partial p}{\partial \rho}.
    \label{temp}
\end{equation}
In the following we will discuss several cases for the matter production rate $\Gamma$.
\\

$\bullet$ $\Gamma = \mbox{constant}$
\\
For a barotropic EoS given in terms of the effective parameter state defined in Eq. (\ref{eq:omegeff}), we obtain in the last expression
\begin{equation}
    \frac{\dot{T}}{T} = \omega_{\mathrm{eff}}\frac{\dot{n}}{n} = -3H\omega_{\mathrm{eff}}\left(1-\frac{\Gamma}{3H} \right),
\end{equation}
where the equation (\ref{conseqs}) was considered. In this case an explicit function of time for the Hubble parameter was given in Eq. (\ref{eq:hubb}) then by direct integration we can write for the temperature
\begin{equation}
    T(t) = T(t_{0})\exp \left[-2A \left(1-\exp \left\lbrace-\frac{\Gamma}{2}(t-t_{0})\right\rbrace \right) \right].
\end{equation}
It is worthy to mention that depending on the values of the constant $A$, $A>0$ ($A<0$), the temperature will have a decreasing (increasing) behavior. In the limit case, $t \rightarrow \infty$, we have $T(t) \rightarrow T(t_{0})\exp(-2A)$, therefore $T(t \rightarrow \infty) < T(t_{0})$ for $A > 0$ and $T(t \rightarrow \infty) > T(t_{0})$ for $A < 0$.
\\

$\bullet$ $\Gamma = 3\beta H_{0}\left(\frac{H}{H_{0}} \right)^{\alpha}$
\\
For this model we will consider first the case $\alpha = 1$, then using the expressions (\ref{besth}), (\ref{bestomega}) and (\ref{temp}), we obtain
\begin{equation}
    T(z) = T_{0}(1+z)^{-3\beta(1-\beta)},
\end{equation}
for $\beta > 1$ (phantom regime) we have decreasing behavior for the temperature as the universe expands. On the other hand, for $ 1/3 < \beta < 1$ (quintessence) the temperature increases and becomes singular at the far future, $z=-1$. For $\alpha \neq 1$, one gets
\begin{equation}
    T(z) = T_{0}\exp \left[\frac{2\beta}{1-\alpha}\left\lbrace \beta + (1-\beta)(1+z)^{3(1-\alpha)/2} \right\rbrace^{-1} \right],
\end{equation}
this temperature has a bounded value at the far future given by $T(z=-1) = T_{0}\exp[2/(1-\alpha)]$, which is independent of the constant $\beta$. Recalling that the interesting case for this model is given by $\alpha < 1$ and $\beta > 1/3$, we have that the temperature given in the previous expression tends to increase as universe evolves.
\\

$\bullet$ $\Gamma = 3H\gamma$
\\
This model was discussed in Refs. \cite{Nunes:2015rea, victor, victor1}, being $\gamma$ a positive constant. By means of Eqs. (\ref{conseqs}) and (\ref{hdteq}) we can write
\begin{equation}
    n(t) = n(t_{0})\left(\frac{a(t)}{a(t_{0})}\right)^{-3(1-\gamma)},
\end{equation}
and the following Hubble parameter
\begin{equation}
    H(t) = H_{0}\left[1+\frac{3}{2}H_{0}(1-\gamma)(t-t_{0}) \right]^{-1}.
\end{equation}
For $\gamma > 1$, the previous expression can be written as
\begin{equation}
H(t) = \frac{2}{3\left|1-\gamma \right|}(t_{s}-t)^{-1},    
\end{equation}
where $t_{s}:= t_{0}+2/(3H_{0}\left|1-\gamma \right|)$, i.e., $t_{s}$ denotes a time in the future at which the Hubble parameter becomes singular, this is characteristic of a phantom scenario, always that the condition, $\gamma > 0$ is satisfied, we will have an expanding universe. In this case the effective parameter state (\ref{goodone}) can be written as, $\omega_{\mathrm{eff}} = -\gamma$. Therefore, depending on the value of the constant $\gamma$, the model admits a quintessence, cosmological constant or phantom scenarios. The expression (\ref{temp}) for the evolution of temperature reads
\begin{equation}
    \frac{\dot{T}}{T} = 3\gamma (1-\gamma)H(t),
\end{equation}
which results after a straightforward integration as
\begin{equation}
    T(t) = T(t_{0})\left[1+\frac{3}{2}H_{0}(1-\gamma)(t-t_{0})\right]^{2\gamma}.
\end{equation}
As the universe expands, the dark matter content warms.
\\

$\bullet$ $\Gamma = \delta /n$
\\
For this case we will consider, $\delta = \mbox{constant}$. From Eq.~(\ref{conseqs}) we can obtain the following expression for the particle number density
\begin{equation}
    n(t) = n(t_{0})\left[1+\frac{\delta}{n(t_{0})}\int^{t}_{t_{0}}\left(\frac{a(t)}{a(t_{0})} \right)^{3}dt \right]\left(\frac{a(t_{0})}{a(t)} \right)^{3},
    \label{ndens}
\end{equation}
if we assume that, $\rho = mn$, (non-relativistic matter) being $m$ the rest mass for the dark matter particle, then from the last equation we can write
\begin{equation}
    \rho(t) = \rho(t_{0})\left[1+\frac{\delta}{n(t_{0})}\int^{t}_{t_{0}}\left(\frac{a(t)}{a(t_{0})} \right)^{3}dt \right]\left(\frac{a(t_{0})}{a(t)} \right)^{3},
\end{equation}
where $\rho_{0} = mn(t_{0})$. From the use of the Friedmann constraint, $3H^{2} = \rho$, it is possible to establish an expression for the Hubble parameter. For this case the effective parameter state takes the form
\begin{equation}
    \omega_{\mathrm{eff}}(t) = -\frac{\delta}{3nH},
    \label{last}
\end{equation}
where we have considered the continuity equation for the energy density as in the previous models. In order to have a phantom scenario at present time the condition, $\delta > 3n_{0}H_{0}$, must be satisfied. Therefore, by means of Eqs. (\ref{conseqs}), (\ref{temp}) and the effective parameter (\ref{last}), we can write
\begin{equation}
    \frac{\dot{T}}{T} = \frac{\delta}{n}\left(1+\omega_{\mathrm{eff}} \right),
\end{equation}
which leads to
\begin{equation}
    T(t) = T(t_{0})\exp \left\lbrace \delta \int^{t}_{t_{0}}\left[\frac{1+\omega_{\mathrm{eff}}(t)}{n(t)} \right] dt \right\rbrace.
\end{equation}
For a phantom scenario, the temperature will have an decreasing behavior since, $1+\omega_{\mathrm{eff}} < 0$. On the other hand, for quintessence the temperature will increase as universe expands.
\\

$\bullet$ $\Gamma = (3\alpha H)/n$
\\

Let us assume, $\alpha = \mbox{constant}$. From the continuity equation for the energy density (\ref{conseqs}), we have
\begin{equation}
    \omega_{\mathrm{eff}}(a)= -\frac{\alpha}{n(a)},
    \label{pofeos}
\end{equation}
in order to have phantom expansion at present time, the condition $\alpha > n_{0}$ must be fulfilled. By integrating the density number equation (\ref{conseqs}), one gets
\begin{eqnarray}
    n &=& \left(\frac{a_{0}}{a}\right)^{3}\left[n_{0}-\alpha+\alpha\left(\frac{a}{a_{0}}\right)^{3} \right]\nonumber,\\
    &=& \alpha + n_{0}(1+\omega_{\mathrm{eff,0}})(1+z)^{3}.
\end{eqnarray}
where $\omega_{\mathrm{eff,0}}$ represents the value of the effective parameter at present time and the Eq. (\ref{pofeos}) was used together with the definition of the redshift given before. Therefore we can observe that for this model, $\omega_{\mathrm{eff}}(z = -1) = -1$. If we integrate the continuity equation for the energy density we can obtain
\begin{equation}
    \rho(z) = \frac{\rho_{0}}{n_{0}}\left\lbrace\alpha +n_{0}(1+\omega_{\mathrm{eff,0}})\left(1+z\right)^{3}\right\rbrace,
\end{equation}
using the Friedmann constraint we can establish the form of the Hubble parameter as a function of the redshift, $H(z) = \sqrt{\rho(z)/3}$. In this case we have, $H(z\rightarrow -1) \rightarrow \sqrt{-(\rho_{0}\ \omega_{\mathrm{eff,0}})/3}$, resulting that Hubble parameter in this model is similar to the $\Lambda$CDM Hubble parameter, i.e., as we approach to the far future the Hubble parameter tends to a bounded value. In this case from Eq. (\ref{temp}) we can obtain for the temperature
\begin{equation}
    T(z) = T_{0}\exp \left\lbrace \omega_{\mathrm{eff,0}}\frac{ (1+\omega_{\mathrm{eff,0}})(1-(1+z)^{3})}{(1+\omega_{\mathrm{eff,0}})(1+z)^{3}-\omega_{\mathrm{eff,0}}}\right\rbrace,
\end{equation}
where the integration was carried out from $0$ to $z$. Note that for a phantom scenario the value of the temperature will increase as the universe expands and for quintessence the universe is cooling down.
\\

It is worthy to mention that all the models we have studied in this section can be expressed through the formula
\begin{equation}
    \Gamma = 3 \beta H_{0} \left(\frac{H}{H_0}\right)^{\alpha}.
\end{equation}
For $\Gamma = \text{constant}$, we can simply set $\alpha = 0$; for the model $\Gamma = 3H\gamma$, we set $\alpha = 1$ and recognize $\beta = \gamma$. The model $\Gamma = 3H\alpha/n$ is equivalent to the case where $\alpha = -1$, because from Friedman equation $n \propto H^2$. Finally, for the same reason our model $\Gamma = \delta /n$ is equivalent to the case where $\alpha = -2$. However, in order to distinguish the thermodynamics characteristics of each model, we have studied each case separately.
\\

In general grounds, the constitution of DM is still an unsolved problem in cosmology and particle physics. However, nowadays several well-motivated DM candidates are under scrutiny. The abundance of DM in our observable universe must have its origin in the early universe and in at least two different forms: thermal and non-thermal production, with this we will refer to processes in equilibrium and outside thermodynamic equilibrium, respectively. One of the possibilities for DM are the so-called WIMPs (Weakly Interactive Massive Particles), which are considered as thermal relics. In the early universe the WIMPs density number, $n_{W}$, is governed by the Boltzmann equation $\dot{n}_{W} = -3Hn_{W} - \langle \sigma v \rangle (n^{2}_{W}-n^{2}_{eq})$, where $\langle \sigma v \rangle$ is the thermally averaged WIMPs annihilation cross section times WIMPs relative velocity and $n_{eq}$ is the equilibrium density, we must note the similarity of the aforementioned equation with Eq. (\ref{conseqs}). Therefore, the WIMPs density at present time is caractherized by $\Omega_{W} \propto 1/\langle \sigma v \rangle$, which gives the correct present day density of DM and a solution for the (thermalized) density number is of the form $n_{W} \propto \exp [-(m_{W}-\mu)/T]$, being $\mu$ the chemical potential and $T$ a constant temperature. Thus, the annihilation cross section together with the temperature are important quantities for the description of WIMPs, a complete and interesting review on this topic can be found in Ref.~\cite{wimps}. On the other hand, as discussed in this section, depending on the specific form of the matter production rate, $\Gamma$, the DM temperature remains positive but can have an increasing (decreasing) behavior, such conduct seems to be in contradiction with the description of particle physics given above. However, recent results show that DM can be heated up and displaced from the center of certain galaxies as a result of stars formation \cite{heat}.        

\section{Observational constraints}
\label{sec:observations}

In this section we test the statistical performance of the models we have presented in Section \ref{sec:close} against recent low redshift data both from type Ia supernova and $H(z)$ measurements.\\

For type Ia supernova we make use of the full Pantheon sample \cite{Scolnic} incorporating the heliocentric redshift, correcting in this way the issue previously mentioned about peculiar velocities at high redshift \cite{sarkar}. The Hubble parameter as a function of redshift $H(z)$ data is obtained by cosmic chronometers and taken from the compilation made in \cite{hdzdata}. The statistical analysis was made using the code EMCEE \cite{emcee}, a Python code of the affine-invariant ensemble sampler for Markov chain Monte Carlo (MCMC) proposed by Goodman and Weare \cite{GW2010}.

\subsection{$\Gamma = \text{constant}$ model}

Let us start with the model $\Gamma = \text{constant}$ discussed in Section \ref{sec:close}. From Eq. (\ref{hdasol}) the Hubble parameter can be penned as
\begin{equation}\label{hdz1}
    H(z) = H_0 \left[ \frac{\Gamma}{3H_0} + (1+z)^{3/2} \left(1 - \frac{\Gamma}{3H_0} \right)\right].
\end{equation}
Written in this way we have two free parameters: $H_0$ and the combination $\Gamma/3H_0$. As starting point of our study we use only type Ia supernova data from the Pantheon sample \cite{Scolnic}. In this case, the observable is the distance modulus
\begin{equation}
    \mu(z)=25+5\log \left( \frac{c}{H_0}(1+z)\int \frac{dx}{E(x)}\right),
\end{equation}
where $E(x):= H(x)/H_0$ is the normalized Hubble parameter. Although in principle the Hubble function has two free parameters to fit, $H_0$ and $\Gamma$, due to the degeneracy of the Hubble constant $H_0$ and the absolute magnitude of the supernovae $M$, $H_0$ is not fixed by the data, and is usually marginalized. After doing that, the only free parameter is the combination $\Gamma/3H_0$.\\

The result of the statistical analysis is $\Gamma/3H_0=0.182 \pm 0.035$. This means; the supernova data alone suggest that $\Gamma >0$, in agreement with our theoretical consideration of section \ref{sec:close}. Also, we get that $\Gamma <3H_0$, that means our time scale of particle creation is larger than the time scale of expansion. We can write then
\begin{equation}\label{gama01}
    \Gamma = (0.182 \pm 0.035)3H_0.
\end{equation}
We also use data from $H(z)$ measurements. This implies the use of Eq. (\ref{hdz1}) directly, in this way we are left only with the parameters $H_0$ and $\Gamma$, free to be constrained. We also add the prior information for the Hubble constant $H_0$. Because there is a well known tension in the value of $H_0$ using different methods, we perform our study using both values informed in \cite{PDG}: $h = 0.678 \pm 0.009$ (Planck) and $h=0.732 \pm 0.017$. Let us consider in this subsection the constraint using only $H(z)$ measurements and the gaussian prior on $H_0$. Using first the Planck value for the Hubble constant, we get $h=0.674 \pm 0.013$ and $\Gamma/3H_0 = -0.20 \pm 0.2$, implying that $\Gamma <0$ at 1 $\sigma$, something at odds with our previous considerations. On the other hand, using the other prior, we get $h = 0.71 \pm 0.02$ and $\Gamma/3H_0 = 0.0 \pm 0.2$. In this case we do not have a conclusive results for the sign of $\Gamma$ at 1 $\sigma$.\\

\begin{figure}[h!]
\includegraphics[width=7.5cm]{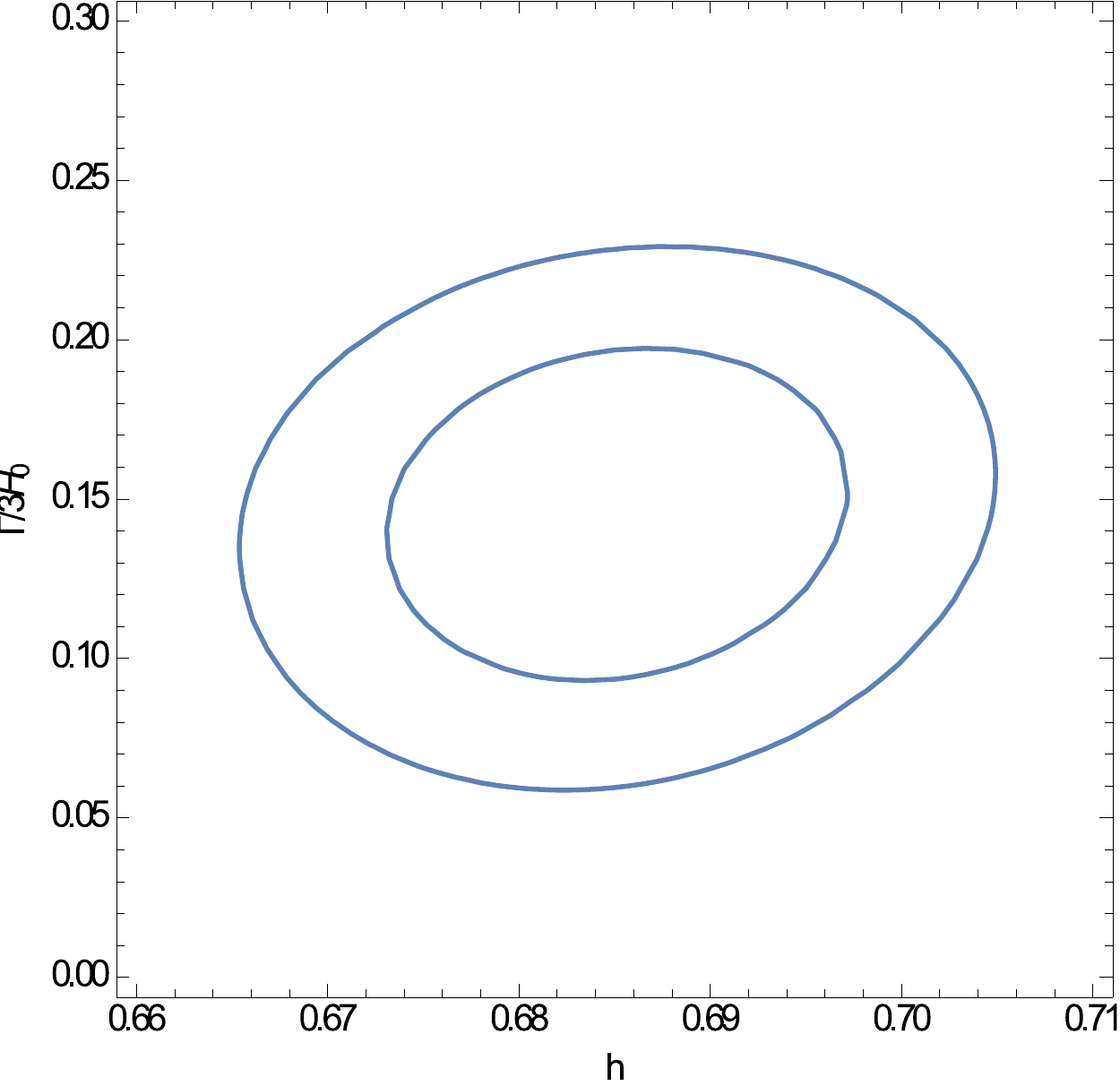}
\caption{We display the results for $ 1\sigma $ and $ 2\sigma $ for our model in the parameter space $(h, \Gamma/3H_0)$ using both data and the Planck prior.}
\label{fig:fig1}
\end{figure}

\begin{figure}[h!]
\includegraphics[width=7.8cm]{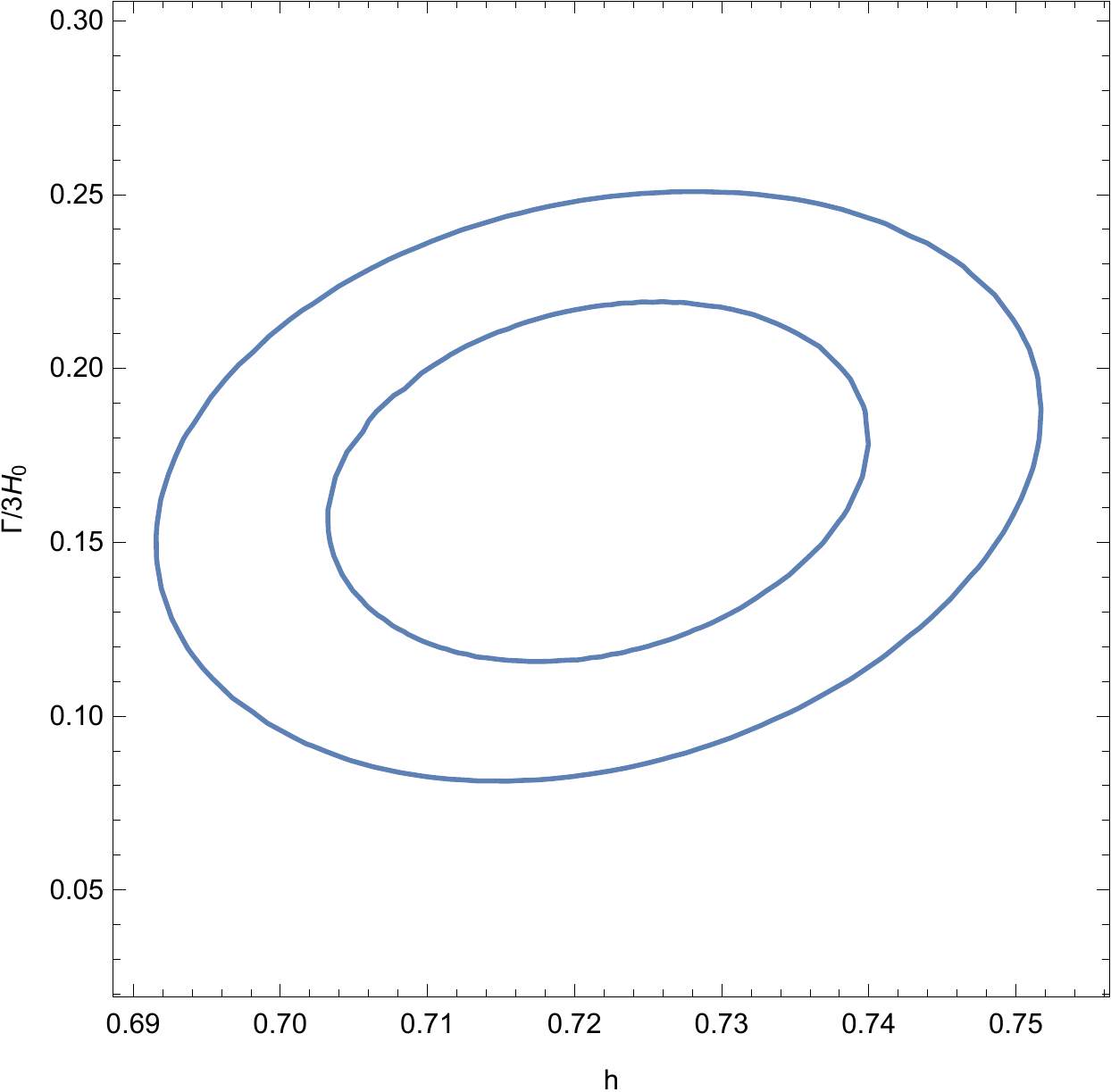}
\caption{We display the results for $ 1\sigma $ and $ 2\sigma $ for our model in the parameter space $(h, \Gamma/3H_0)$ using both data and the second prior for $H_0$.}
\label{fig:fig2}
\end{figure}

Let us study the results using both observational probes together. Using the Gaussian prior from Planck, the best fit values are:  $h=0.685 \pm 0.010$ and $\Gamma/3H_0 = 0.15 \pm 0.05$. The confidence contours are displayed in Fig. (\ref{fig:fig1}). From this analysis, the best value is $\Gamma = 30 \pm 9$, which is similar to the previous estimation (\ref{gama01}), that gives the value $\Gamma = 36 \pm 6$. By repeating the analysis but now using the second prior for the Hubble constant we get $h=0.72 \pm 0.02$ and $\Gamma/3H_0 = 0.168 \pm 0.045$. The confidence contours are displayed in Fig. (\ref{fig:fig2}).

\subsection{Testing the best matter creation model}

\begin{figure}[h!]
\includegraphics[width=9cm]{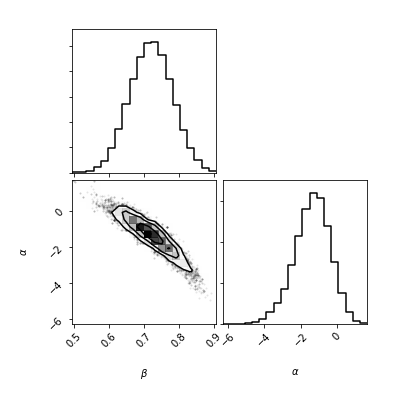}
\caption{We display the results for $ 1\sigma $, $ 2\sigma $ and $3 \sigma$ for our model in the parameter space $(\alpha , \beta )$ using only data from type Ia supernova.} 
\label{fig:fig3}
\end{figure}

In what follows, we perform a statistical test using the model \cite{generalrate, generalrate1} described by the particle creation rate
\begin{equation}
    \Gamma = 3\beta H_{0}\left(\frac{H}{H_{0}}\right)^{\alpha}.
\end{equation}
As we described in section \ref{sec:close}, by inserting the previous expression in the conservation equation and then solving for the Friedmann equation, we get a differential equation for the Hubble function $H(z)$, whose solution for $\alpha \neq 1$ is
\begin{equation}
    H(z)=H_0 \left[ \beta + (1-\beta)(1+z)^{3(1-\alpha)/2} \right]^{1/(1-\alpha)}.
\end{equation}
Although in doing this we can not assess the qualities of each model discussed in Section \ref{sec:temp}, we can use the test to find the best fit values for $\alpha$ and $\beta$, and after that we can discover which model is certainly closer to the best one suggested by observations. The statistical analysis using only the type Ia supernova data gives the following best fit values, $\alpha = -1.3 \pm 0.8$ and $\beta = 0.72 \pm 0.06$. In Figure (\ref{fig:fig3}) we show the triangle plot showing the posterior probability for each parameter and the contour plot.

\begin{figure}[h!]
\includegraphics[width=9cm]{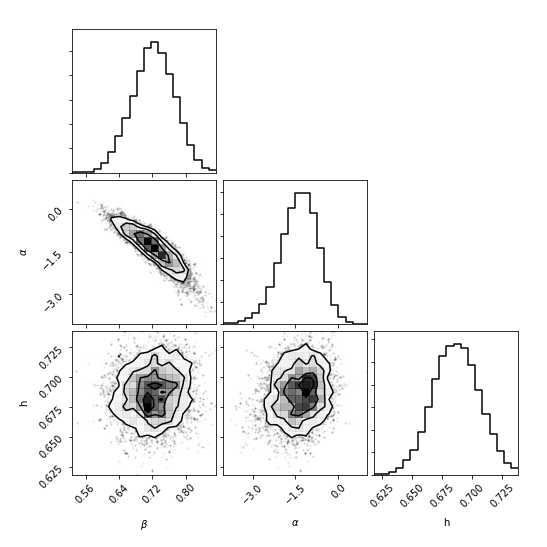}
\caption{We display the results for $ 1\sigma $, $ 2\sigma $ and $3 \sigma$ for our model in the parameter space $(\alpha , \beta, h )$ using both data from type Ia supernova and $H(z)$.}
\label{fig: fig4}
\end{figure}

These results are not surprising, because at 1 $\sigma$ confidence level the CCDM solution it is contained. This solution corresponds to a matter creation model completely analogous to the $\Lambda$CDM solution for which $\alpha = -1$ and $\beta = \Omega_{\Lambda} \simeq 0.7$ \cite{generalrate, generalrate1}. Recall that $h$ has been marginalized together with the maximum absolute magnitude for the supernovas. Once we add the $H(z)$ data, we have all three parameters free to constraint, $\alpha$, $\beta$ and $h$. The result of our MCMC analysis is display in figure (\ref{fig: fig4}).

The best fit values of the parameters are $\beta = 0.72 \pm 0.05$, $\alpha = -1.3 \pm  0.5$, and $h = 0.69 \pm 0.018$. Again, the CCDM model seems to be preferred in contrast to the other models.

\section{Incorporating other effects}
\label{sec:other}

We now consider the introduction of cosmological constant, $\Lambda$, in the framework of matter creation models. This consideration only modifies the Friedmann constraint as $3H^{2} = \rho + \Lambda$, therefore the Eq. (\ref{hdteq}) can we re-expressed in the following form
\begin{equation}
    \frac{\dot{H}}{H^{2}} = -\frac{3}{2}\left(1-\frac{\Lambda}{3H^{2}}\right)\left(1-\frac{\Gamma}{3H}\right),
\end{equation}
which can be written in terms of the deceleration parameter straightforwardly. If we consider the case $\Gamma = \mbox{constant}$ and evaluate at present time the previous equation one gets
\begin{equation}
    q_{0} = -1 + \frac{3}{2}\left(1-\Omega_{\Lambda,0}\right)\left(1-\frac{\Gamma}{3H_{0}}\right).
\end{equation}
In the $\Lambda$CDM model the deceleration parameter is given by the expression $q(z) = - 1 + 3/(2[1+\frac{\Omega_{\Lambda,0}}{\Omega_{m,0}}(1+z)^{-3}])$ and the normalization condition $\Omega_{\Lambda,0} + \Omega_{m,0} =1$ must be satisfied. According to the last Planck collaboration results, $\Omega_{m,0} = 0.315 \pm 0.007$ \cite{plancknew}, therefore the deceleration parameter lies in the interval $-0.538 \leq q_{0} \leq -0.517$, using these results we are left with the condition
\begin{equation}
     0.043 \leq \frac{\Gamma}{3H_{0}} \leq 0.045,
\end{equation}
then, by means of Eq. (\ref{goodone}) we can not have a transient phantom behavior at present time. We obtain the same interval for the constant $\beta$ if we consider the model $\Gamma = 3\beta H_{0}(H/H_{0})^{\alpha}$.\\

As mentioned before, if we set $\alpha = 1$ we can recognize $\gamma = \beta$, being $\gamma = \mbox{constant}$ in the model $\Gamma = 3H\gamma$, which provides $\omega_{\mathrm{eff}} = -\gamma$. Therefore, the inclusion of cosmological constant does not provide the possibility of crossing the phantom divide in this model. We do not share the idea of authors of Ref. \cite{Nunes:2015rea}, where was claimed that a matter production rate given as $\Gamma = 3H\gamma$ plus a cosmological constant can give a phantom scenario. On the other hand, if we neglect the cosmological constant contribution in the aforementioned model, the crossing to the phantom divide is possible only if $\gamma > 1$, but, the statistical analysis revealed that $\gamma < 1$.\\

A second possibility to consider is given by the inclusion of bulk viscous effects, for such models have been shown that at effective level can have a phantom (quintessence) behavior with no need of extra ingredients, see for instance Refs. \cite{viscous1, viscous2, CLO18, CCL17}, in this case the Friedmann constraint and acceleration equation are given by
\begin{equation}
    3H^{2} = \rho, \ \ \ \dot{H}+H^{2} = -\frac{1}{6}\left[\rho + 3(p+\Pi) \right],
\end{equation}
being $p$ the local equilibrium pressure, $\Pi < 0$ is the bulk viscous pressure and the continuity equation for energy density reads $\dot{\rho}+3H(\rho + p + \Pi) =0$, which provides the following effective parameter of state
\begin{equation}
    \omega_{\mathrm{eff}}(t) = \frac{p(t)+\Pi(t)}{\rho(t)}.
\end{equation}
the most simple definition for the viscous pressure is given by $\Pi = -3\xi(\rho) H$, where $\xi$ is the bulk viscous coefficient. However, this election for $\Pi$ leads to non causality and it is well known as Eckart model. On the other hand, in the Israel-Stewart model the viscous pressure must obey a transport differential equation and in such case the theory respects causality. Using the continuity equation given above and the expression given in (\ref{eq:contmod}) one gets
\begin{equation}
    -\frac{\Gamma}{3H} = \frac{\Pi}{p+\rho},
\end{equation}
for a pressureless fluid we can write
\begin{equation}
    -\frac{\Gamma}{3H} = \frac{\Pi}{\rho} = \omega_{\mathrm{eff}}.
\end{equation}
Note that despite the inclusion of dissipative effects, the effective parameter state has the same form of equation (\ref{goodone}), in addition, regardless of the choice we make for $\Pi$, i.e., the one given by the Eckart model or the solution arising in the tranport equation within the description of Israel-Stewart; both theories describe dissipative processes near equilibrium, this condition is given by
\begin{equation}
    \left|\frac{\Pi}{\rho} \right| \ll 1,
\end{equation}
therefore the effective parameter state will not cross the phantom divide.

\section{Inhomogeneous matter production rate}
\label{sec:inhomo}
In this section we explore other possibility for the matter production rate, $\Gamma$. In general, matter can couple with the curvatures, which include $\dot H$ in addition to $H$. Then it might be more natural to assume $\Gamma$ could also depend on $\dot H$, 
\begin{equation}
\label{Gamma1}
\Gamma = \Gamma \left( H, \dot H \right) \, .
\end{equation}
or in more general, 
\begin{equation}
\label{Gamma2}
\Gamma = \Gamma \left( \rho, p, H, \dot H, \ddot H, \dddot H, \cdots \right) \, .
\end{equation}
For usual perfect fluid $p$ is given by an equation of state, $p\left(\rho\right)$. Thus we obtain from the continuity equation (\ref{conseqs})
\begin{eqnarray}
\label{Gamma3}
\dot \rho &+& 3 H \left( \rho + p \left(\rho\right) \right) - \nonumber \\ 
&-&  \left( \rho + p \right)
\Gamma \left( \rho, p\left(\rho\right), H, \dot H, \ddot H, \dddot H, \cdots \right) =0 \, ,
\end{eqnarray}
from which we can write 
\begin{eqnarray}
\label{Gamma4}
\Pi &=& \Pi \left( \rho, p\left(\rho\right), H, \dot H, \ddot H, \dddot H, \cdots \right) \nonumber \\
&=&  - \frac{\rho + p}{3H} 
\Gamma \left( \rho, p\left(\rho\right), H, \dot H, \ddot H, \dddot H, \cdots \right) \, ,
\end{eqnarray}
which may be regarded as a kind of the generalized equation of state proposed in \cite{Nojiri:2005sr} because $\Pi$ is an effective pressure. Instead of the standard equation of state $p=p\left(\rho\right)$, we may also consider the general equation of state in \cite{Nojiri:2005sr} as follows, 
\begin{equation}
\label{Gamma4B}
p = p \left( \rho, H, \dot H, \ddot H, \dddot H, \cdots \right) \, .
\end{equation}
In (\ref{Gamma3}), $\rho$ and $p$ include both of the contributions from the matter and the cosmological constant. Just for the illustrative reasons, we only consider the contribution from matter, that is, we assume $p=0$. Then for a simple model, 
\begin{equation}
\label{Gamma5}
\Gamma = \gamma' \left(H\right) \dot H \, ,
\end{equation}
with a function $\gamma \left(H\right)$. Then Eq.~(\ref{Gamma3}) can be solved as 
\begin{equation}
\label{Gamma6}
\rho = \rho_{0} a^{-3} e^{\gamma \left(H\right)} \, ,
\end{equation}
with a constant of the integration $\rho_0$. Then the first FLRW equation has the following form 
\begin{equation}
\label{Gamma7}
3 H^2 = \rho_{0} a^{-3} e^{\gamma \left(H\right)} \, .
\end{equation}
In case 
\begin{equation}
\label{Gamma8}
\gamma \left(H\right) = \left( 1 - \frac{1}{\beta} \right) 
\ln \left(3 H^2 \right) \, ,
\end{equation}
with a constant $\beta$, the first FLRW equation (\ref{Gamma7}) can be rewritten as 
\begin{equation}
\label{Gamma9}
3 H^2 = \rho_{\mathrm{eff}} \, , \quad 
\rho_{\mathrm{eff}} \equiv \rho_0^\beta a^{-3 \beta} \, .
\end{equation}
Because the energy density of the perfect fluid with the constant equation of the state parameter behaves as $a^{-3\left( 1 + \omega \right)}$, Eq.~(\ref{Gamma9}) tells that the effective equation of state parameter $\omega_\mathrm{eff}$ is given by 
\begin{equation}
\label{Gamma10}
\omega_{\mathrm{eff}} = -1 + \beta \, .
\end{equation}
Therefore if $\beta$ is negative, there appears the effective phantom where $\omega_{\mathrm{eff}} < -1$ and if $0<\beta<\frac{2}{3}$, there appears the effective quintessence, $-\frac{1}{3} > \omega_{\mathrm{eff}} > -1$.  General development of the expansion in the universe might be realized by using more complex function $\gamma \left(H\right)$. As another example, we may consider 
\begin{equation}
\label{Gamma11}
\gamma(H) = \ln \left( \frac{H^2}{H_0^2} \right) 
 - \frac{9}{4} \ln \left( \frac{H^2}{H_0^2} - 1 \right) \, ,
\end{equation}
with a constant $H_0$. Then the solution of (\ref{Gamma7}) is given by
\begin{align}
\label{Gamma12}
& a(t) = A \sinh^\frac{3}{2} \left( \frac{2H_0}{3} t \right) \, , \quad H(t) = H_0 \coth \left( \frac{2H_0}{3} t \right) \, , \nonumber \\ 
& A \equiv \left( \frac{\rho_0}{3 H_0^2} \right)^\frac{1}{3} \, .  
\end{align}
The solution in (\ref{Gamma12}) is identical with that of the $\Lambda$CDM model. In our model, originally there is only matter, which may be identified with the cold DM, but by the effect of the particle creation, there appears the effective cosmological constant. We should note that the de Sitter space-time, that is, $H=H_0$ $\left(H_0:\mbox{constant}\right)$, $a(t)\propto e^{H_0 t}$ is not a solution of (\ref{Gamma7}) for any choice of $\gamma\left( H \right)$ because the l.h.s. of (\ref{Gamma7}) is constant but the r.h.s. changes in time as $\propto e^{-3 H_0 t}$ for arbitrary $\gamma\left( H \right)$ . We now consider general case, except the pure de Sitter space-time, that $H$ depend on time $t$, $H=H(t)$, which can be assumed to be solved with respect to $H$ as $t=t(H)$. Then, (\ref{Gamma7}) tells that the function $\gamma \left(H\right)$ is explicitly given by a function of $H$ as follows, 
\begin{equation}
\label{Gamma13}
\gamma \left(H\right) = \ln \left( \frac{3 H^2}{\rho_0} a\left( t\left(H\right) 
\right)^3 \right) \, .
\end{equation}
Then for the arbitrary evolution of the scale factor $a=a(t)$, except the pure de Sitter space-time, because $H(t)=\frac{\dot a(t)}{a(t)}$, the evolution can be realized by choosing $\gamma \left(H\right)$ by (\ref{Gamma13}). Therefore, for example, we can construct models which unifies the inflation in the early universe and the late-time accelerating expansion in the dark energy era. 
 
\section{Final remarks}
\label{sec:final}

In this work we have studied some cosmological aspects of matter creation models and their ability to represent a phantom regime at effective level. We considered two cases for the matter production rate, $\Gamma$; the simplest election is given by a constant production rate and as second choice for this term we took into account a $\Gamma$-term given as a power-law of the Hubble parameter \cite{generalrate, generalrate1}. However, we provided some other examples for $\Gamma$ in order to discern if some differences can be found in the thermodynamics description of each model. Since no other contribution was considered on the cosmic fluid, we have that this type of model it is able to describe the universe at late times. Additionally, the matter created characterizes DM since we have setted its pressure equal to zero.\\ 

By considering the case in which the $\Gamma$-term is given by a constant it was possible to identify that the energy density of the fluid it is composed by three terms: dark matter, cosmological constant and a third term that could characterize a fluid with parameter state, $\omega < 0$. Specifically, this election leads to a transient phantom or quintessence evolution depending on the values of the parameters involved. On the other hand, this transient behavior for the phantom or quintessence scenarios was also obtained for the general $\Gamma$-term considered in this work. It is worthy to mention that this transitory scheme can be found in some other cosmological models where the main interest was to devise an universe free of future singularities. Besides, a main characteristic found in all the models discussed in this work is that they contain a de Sitter expansion, i.e., depending on the $\Gamma$-term, the model can evolve from quintessence or phantom regimes at present time to a cosmological constant like expansion as we approach the far future or can imitate a cosmological constant throughout cosmic evolution; this last characteristic it is obtained only in one of the examples provided for the matter production rate term, where the effective parameter state is given by a constant.\\

The matter production models are beyond the standard cosmological model since the cosmic expansion it is not adiabatic, i.e., the matter creation effects contribute to the generation of entropy \cite{maartens}, which is a more consistent description. Therefore, from the thermodynamics point of view, the temperature associated to the created matter will evolve as the universe expands. As we found in this work, according to the election of the matter production rate, the temperature will have a decreasing or increasing role as universe unfolds but keeps positive. This last conduct seems to be in agreement with some recent observations \cite{heat}, where DM can be heated up due to the formation of stars in some galaxies. The temperature could also become singular in the far future, of course this depends on the elected $\Gamma$-term.\\ 

However, despite all the interesting features that may be associated with cosmological matter creation models, these are not favored by observations if the intention is to describe an effective phantom regime at present time. In general, for these models the effective parameter state is given by the expression
\begin{equation*}
    \omega_{\mathrm{eff}} = -\frac{\Gamma}{3H}.
\end{equation*}
Notice that depending on the election of the $\Gamma$-term, the effective parameter state can vary. This expression holds under the incorporation of other effects in the cosmological fluid; cosmological constant for example. For the case $\Gamma = \mbox{constant}$, the model is allowed to cross the phantom divide always that the condition, $\Gamma > 3H_{0}$, is fulfilled. However, according to the observational analysis, the quotient $\Gamma/3H_{0}$ is always less than 1. This result implies that if the DM sector it is supported by a particle description (WIMPs for instance); such particles never reach the thermal equilibrium \cite{wimps}. On the other hand, if we focus on the general case for the $\Gamma$-term, we can see that $\omega_{\mathrm{eff,0}} = -\beta$, therefore from the results obtained in the observational analysis; the value constrained for the constant $\beta$ it is compatible only with a quintessence scenario. It is worthy to mention that the lower bounds obtained for the value constrained for the constant $\beta$ with the use of observational data are in good agreement with the upper bound obtained for the parameter state of dynamical DE models in Ref.~\cite{des}, where $\omega_{\mathrm{de,0}} = -0.95^{+0.33}_{-0.39}$.\\ 

On the other hand, given the positivity of the temperature, the resulting quintessence DE scenarios in this approach will also have positive entropy given that the Euler relation establishes that the product of both quantities is proportional to $(1+\omega)$, i.e., the accelerating universe in this description will not have the negative entropy or negative temperature problem \cite{CLO18}.\\ 

In conclusion, this class of models can not cross the phantom divide even if we include a cosmological constant or some other effects in the cosmological fluid such as bulk viscosity. However, if we consider an inhomogeneous expression for the $\Gamma$-term as discussed in section \ref{sec:inhomo}, we can observe that such election leads to an effective phantom/quintessence behavior, notice that this consideration is in fact a generalization for the models considered in this work since also derivatives of the Hubble parameter are allowed. We will discuss in detail this kind of model of matter creation elsewhere.\\ 

Our results differ from those obtained in Ref. \cite{Nunes:2015rea}, where was stated that creation models plus a cosmological constant can describe a phantom scenario at effective level. Finally, as commented previously, the constant $\beta$ can be related directly with the parameters involved in each case discussed for the $\Gamma$-term in section \ref{sec:temp} of this work, therefore the exclusion from the phantom regime applies for all the cases considered here.\\   

\section*{Acknowledgments}
This work has been supported by S.N.I. (CONACyT-M\'exico) (M.C.) and by MINECO (Spain), FIS2016-76363-P, and by project 2017 SGR247 (AGAUR, Catalonia) (S.D.O). This work is also supported by MEXT KAKENHI Grant-in-Aid for Scientific Research on Innovative Areas ``Cosmic Acceleration'' No. 15H05890 (S.N.) and the JSPS Grant-in-Aid for Scientific Research (C) No. 18K03615 (S.N.).


\begin{references}
\bibitem{obs}
P.~A.~R. Ade, et~al. (Planck Collaboration), Astron.\ Astrophys. {\bf 571}, A16 (2014); Astron.\ Astrophys.
{\bf 594}, A13 (2016); A.~Rest, et~al. Astrophys.\ J. {\bf 795}, 44 (2014).

\bibitem{plancknew}
N.~Aghanim, et~al. (Planck Collaboration), arXiv:1807.06209 [astro-ph.CO].
 
\bibitem{dec1}
R.~R.~Caldwell, Phys.\ Lett.\ B {\bf 545}, 23 (2002). 

\bibitem{dec2}
S.~M.~Carroll, M.~Hoffman and M.~Trodden, Phys.\ Rev.\ D {\bf 68}, 023509 (2003).

\bibitem{odintsovcft}
S.~Nojiri and S.~D.~Odintsov, Phys.\ Lett.\ B {\bf 562}, 147 (2003).

\bibitem{gibbons}
G.~W.~Gibbons, arXiv:hep-th/0302199.

\bibitem{prl}
R.~R.~Caldwell, M.~Kamionkowski and N.~N.~Weinberg, Phys.\ Rev.\ Lett. {\bf 91}, 071301 (2003).

\bibitem{cft1} 
  E.~Elizalde, S.~Nojiri and S.~D.~Odintsov,
  Phys.\ Rev.\ D {\bf 70}, 043539 (2004).

\bibitem{cft2} 
  S.~Nojiri and S.~D.~Odintsov,
  Phys.\ Rev.\ D {\bf 70}, 103522 (2004).

\bibitem{Nunes:2015rea} 
  R.~C.~Nunes and D.~Pav\'on,
  Phys.\ Rev.\ D {\bf 91}, 063526 (2015).
  
\bibitem{Nojiri:2005sr} 
  S.~Nojiri and S.~D.~Odintsov,
  Phys.\ Rev.\ D {\bf 72}, 023003 (2005).
  
\bibitem{lima}
L.~R.~W.~Abramo and J.~A.~S.~Lima, Class.\ Quantum\ Grav. {\bf 13}, 2953 (1996). 

\bibitem{navarrete}
M.~Cruz, S.~Lepe and G.~Morales-Navarrete, Class.\ Quantum\ Grav. {\bf 36}, 225007 (2019).  
  
\bibitem{victor}
V.~H.~C\'ardenas,
  Eur.\ Phys.\ J.\ C {\bf 72}, 2149 (2012).

\bibitem{victor1}
V.~H.~C\'ardenas, arXiv:0812.3865[astro-ph].

\bibitem{transient1}
N.~Cruz, S.~Lepe and F.~Pe\~na, Eur.\ Phys.\ J.\ C {\bf 72}, 2162 (2012).

\bibitem{transient2}
K.~Dimopoulos, Phys.\ Lett.\ B {\bf 785}, 132 (2018).

\bibitem{generalrate}
M.~P.~Freaza, R.~S.~de~Souza and I.~Waga, Phys.\ Rev.\ D {\bf 66}, 103502 (2002).

\bibitem{generalrate1}
R.~O.~Ramos, M.~Vargas~dos~Santos and I.~Waga, Phys.\ Rev.\ D {\bf 89}, 083524 (2014).

\bibitem{maartens}
R.~Maartens, arXiv:astro-ph/9609119.

\bibitem{wimps}
H.~Baer, Ki-Young~Choi, J.~E.~Kim and L.~Roszkowski, Phys.\ Rep. {\bf 555}, 1 (2015).

\bibitem{heat}
J.~I.~Read, M.~G.~Walker and P.~Steger, Mon.\ Not.\ R.\ Astron.\ Soc. {\bf 484}, 1401 (2019).

\bibitem{Scolnic} 
  D.~M.~Scolnic {\it et al.},
  Astrophys.\ J.\  {\bf 859}, 101 (2018).
  
\bibitem{sarkar} 
  J.~Colin, R.~Mohayaee, M.~Rameez and S.~Sarkar,
  Astron.\ Astrophys.\  {\bf 631}, L13 (2019).
  
\bibitem{hdzdata} 
   A.~G\'omez-Valent and L.~Amendola,
  J.\ Cosmol.\ Astropart.\ Phys. {\bf 1804}, 051 (2018).
  
\bibitem{emcee} 
D.~Foreman-Mackey, D.~W.~Hogg, D.~Lang, and J.~Goodman, Publications of the Astronomical Society of the Pacific, {\bf 125}, 306 (2013).

\bibitem{GW2010}
J.~Goodman and J.~Weare, Commun.\ Appl.\ Math.\ Comput.\ Sci. {\bf 5}, 65 (2010).

\bibitem{PDG}
M.~Tanabashi. et~al. (Particle Data Group), Phys.\ Rev.\ D {\bf 98}, 030001 (2018).
 
\bibitem{viscous1}
N.~Cruz and S.~Lepe, Phys.\ Lett.\ B {\bf 767}, 103 (2017).

\bibitem{viscous2}
M.~Cruz, N.~Cruz and S.~Lepe, Phys.\ Lett.\ B {\bf 769}, 159 (2017).

\bibitem{CLO18}
M.~Cruz, S.~Lepe and S.~D.~Odintsov, Phys.\ Rev.\ D {\bf 98}, 083515 (2018).

\bibitem{CCL17}
M.~Cruz, N.~Cruz and S.~Lepe, Phys.\ Rev.\ D {\bf 96}, 124020 (2017). 


\bibitem{des}
M.~A.~Troxel, et~al. (Dark Energy Survey Collaboration), Phys.\ Rev.\ D {\bf 98}, 043528 (2018).
\end{references}
\end{document}